\documentclass[aps,preprint]{revtex4}

\usepackage{graphicx}
\usepackage{dcolumn}
\usepackage{bm}
\usepackage{color}
\usepackage{epsfig}

\begin{document}

\title{
Spectroscopic properties of large open quantum-chaotic cavities \\ 
with and without  separated time scales 
}
\author{Evgeny N. Bulgakov$^{1,2}$ and Ingrid Rotter$^2$}
\affiliation{$^1$ Kirensky
Institute of Physics, 660036, Krasnoyarsk, Russia}
\affiliation{$^2$ Max Planck Institute for the Physics of Complex Systems,
D-01187 Dresden, Germany}

\date{\today}

\begin{abstract}
The spectroscopic properties of an open large Bunimovich cavity are studied
numerically in the framework of the effective Hamiltonian formalism. The
cavity is opened by attaching two leads to it in four different ways. 
In some cases, 
the transmission takes place via standing waves with an intensity 
that closely follows the profile of the resonances. 
In other cases, 
short-lived and long-lived resonance states coexist. The 
short-lived states cause traveling waves in the transmission 
while the long-lived ones generate superposed fluctuations. 
The traveling waves oscillate as a function of energy. 
They are not localized in the interior of the large chaotic cavity.
In all considered cases, the phase rigidity fluctuates with energy. 
It is mostly near to its maximum value and agrees well with
the theoretical value for the two-channel case. 

\end{abstract}


\maketitle

\section{Introduction}
\label{intro}

During last years, much interest is devoted  to the study of the
spectral properties of small cavities coupled to a small number of channels. 
As a function of the coupling strength between
cavity and attached leads, the results obtained show the resonance trapping
effect in theoretical \cite{seba} as well as in experimental \cite{stm}
studies. Short-lived whispering gallery modes are formed 
by attaching the leads in a suitable manner to  cavities of different shape 
with convex boundary \cite{naz}. These direct processes can result in 
deterministic
transport as signified by a striking system-specific suppression of shot noise
\cite{shot}. The corresponding pathways are localized inside the cavities: 
the whispering gallery
modes near to the convex boundary and the bouncing ball modes around  the
shortest pathway between the two attached leads \cite{naz}.

The whispering gallery modes can be characterized well by classical values 
such as pathway length and travelling time \cite{naz}. The last value is
related to the lifetime of the corresponding state, for example to 
the lifetime of the whispering gallery mode. That means the shorter the
pathways of the direct modes, the shorter the lifetimes of the corresponding 
resonance states and the better these states are separated 
from  the other resonance states of the system by their lifetimes. 
It follows immediately from these relations that it is easy to identify
the whispering gallery modes inside a small system, i.e. in a cavity whose
attached leads have a width that is relatively large as 
compared to the area of the cavity.  It might however be difficult
to identify them in a large system that is relatively weakly coupled to the
leads due to its large area and relatively small width of the attached leads. 
Here the pathway between the two attached
leads is long. The question arises therefore whether or not whispering gallery
modes can be identified also in large cavities.

Guided by the results of previous calculations 
for small cavities \cite{naz,shot}, we have chosen 
billiards of Bunimovich type for the study of this question. We opened these
billiards by attaching two leads to them and used four different 
geometries for 
the positions and orientations of the leads with the aim to study
the transmission  through the cavity under different conditions.
The spectroscopic study is based on the method of the effective 
non-hermitian Hamilton operator $H_{\rm eff}$
that describes the spectroscopic properties of an open quantum system,
i.e. of a quantum system that is opened by embedding it into a common 
continuum of scattering wave functions \cite{rep}.
This means in the present case that leads are attached to the closed cavity
for the propagation of the scattering wave functions
corresponding to one channel in each lead. In some of the 
open cavities, the expectation value $\langle t(E)\rangle$  
of the transmission amplitude $t(E)$ shows an oscillatory behaviour. 
The oscillation length depends on  the geometry of the attached leads. 
It may be large for bouncing ball modes when the distance
between the input and output leads is small. 
It is however much smaller in the case of the whispering gallery modes
that appear in a small attached half stadium and 
are characterized by a relatively 
large distance between the input and output leads. 

The paper is organized in the following manner. In Sec. II, we provide
a few of the basic equations that describe the  different time 
scales in open quantum systems. The time scales are determined by the
lifetimes of the resonance states which are obtained from the eigenvalues of
the effective Hamiltonian $H_{\rm eff}$ that describes the open quantum 
system. They  depend on the manner the
leads are attached to the quantum billiard. We show further
the results of numerical simulations
for the transmission through the four different cavities.
In the case of whispering gallery modes and bouncing ball modes
between the two attached leads, $\langle t(E) \rangle $
oscillates with a period that is determined by the momentum 
$k$ and the geometry of the open cavity.  
In Sec. III, we consider the eigenfunctions of the effective Hamiltonian.
The eigenfunctions are biorthogonal with the consequence that,
in the regime of overlapping resonances, the
real and imaginary parts of the eigenfunctions may decouple,
to a certain degree. We relate this decoupling to the 
phase rigidity of the scattering wave function which expresses the distortion
of the scattering wave function by the overlapping of the different resonance
states. As for the transmission, we calculate the expectation value
of the phase rigidity.  For isolated resonances, 
the real and imaginary parts of the eigenfunctions  are related to one another 
in the standard manner and  the transmission takes place by standing waves. 
In the overlapping regime, however, the  real and imaginary parts of the
eigenfunctions decouple from one another and eventually traveling waves arise.
When fully evolved, the traveling waves are superposed by 
long-lived standing waves.
The traveling waves are described by the so-called optical $S$ matrix. 
The results are summarized in the last section.


\section{Eigenvalues of the effective Hamiltonian and
separated time scales }
\label{eigenv}

\subsection{Eigenvalues and transmission}

The energies and widths of the resonance states of an
open quantum system can be
obtained from the poles of the $S$ matrix or directly from the
eigenvalues $z_k$ of the corresponding effective 
Hamilton operator $H_{\rm eff}$ \cite{rep}. For a quantum billiard
with two attached leads, the effective Hamiltonian is \cite{saro} 
\begin{equation}
H_{\rm eff}=H_B+ \sum_{C=R,L}
V_{BC}\frac{1}{E^{+}-H_C}V_{CB}
\label{Heff}
\end{equation}
where $H_B$ is the Hamiltonian of the closed quantum billiard, $H_C$ is the
Hamiltonian of the left  ($C=L$) and right ($C=R$) lead and
$E^{+}=E+i0$. The second term of $H_{\rm eff}$ takes into account
the coupling of the {\it eigenstates} of $H_B$ via the modes 
propagating in the leads 
when the system is opened. It introduces  correlations between the
states of an open quantum system which appear {\it additionally} to those
of the closed system  \cite{rep}. 
The (real) eigenvalues $E_\lambda^B$ of the Hamiltonian $H_B$ are 
the energies of the discrete states of the closed system, while the 
(complex) eigenvalues $z_\lambda$ of $H_{\rm eff}$ provide  
the positions $E_\lambda$ and  widths $\Gamma_\lambda$ of the corresponding 
resonance states of the 
open system. There is a one-to-one correspondence between the 
number of eigenstates of $H_B$ and that of $H_{\rm eff}$.

Since the effective Hamiltonian (\ref{Heff}) depends explicitely  on 
the energy $E$, so do its eigenvalues $z_\lambda$. The energy dependence is
small, as a rule, in an energy interval that is determined by the 
width of the resonance state. 
The solutions of the fixed-point equations
\begin{eqnarray}
E_\lambda={\rm Re}(z_\lambda)_{|E=E_\lambda}
\label{fixed1}
\end{eqnarray}
and of
\begin{eqnarray}
\Gamma_\lambda=-2\, {\rm Im}(z_\lambda)_{|E=E_\lambda}
\label{fixed2}
\end{eqnarray}
are numbers that coincide approximately with the poles of the $S$ matrix. 
The width $\Gamma_\lambda$ determines
the time scale characteristic of the resonance state $\lambda$. 
The amplitude for the transmission in the one-channel case
is \cite{saro}
\begin{equation}
\label{trHeff}
t=-2\pi
i\sum_{\lambda}\frac{\langle \xi^E_L|V|\phi_\lambda)
(\phi_\lambda|V|\xi^E_R\rangle }{E-z_{\lambda}}  
\end{equation}
where the  eigenfunctions  of $H_{\rm eff}$ are denoted by $\phi_\lambda $
and the scattering wave functions in the leads by $\xi^E_C$.
According to (\ref{trHeff}), the transmission is {\it resonant} in relation 
to the effective Hamiltonian $H_{\rm eff}$. This holds for narrow resonance
states as well as for the short-lived and
long-lived resonance states that appear after redistribution of the 
spectroscopic properties of the system.
Such a redistribution is studied first numerically in a nucleus 
\cite{kleinw} and then analytically, by using statistical assumptions and 
neglecting the energy dependence of the $z_\lambda$, in a large
chaotic system  \cite{sokolov}. It is caused by 
branch points in the complex energy plane \cite{rstrans}.
That means, the eigenvalues 
$z_\lambda$ of $H_{\rm eff}$ determine the time scale of the transmission.
 
The coupling matrix elements $V_{BC}, ~V_{CB}$
between billiard and attached leads can be
calculated in the tight-binding approach \cite{saro,saburo}. 
When  they are small, it is $ E_\lambda \approx E_\lambda^0$ 
and $\Gamma_\lambda \approx \Gamma_\lambda^0$
where $E_\lambda^0$ is the position of the isolated resonance state and 
$\Gamma_\lambda^0$  its width that is determined  by the  $V_{BC}, ~V_{CB}$. 
For large
$V_{BC}, ~V_{CB}$, however,  $E_\lambda^0$ and $ E_\lambda$
as well as $\Gamma_\lambda^0$ and $\Gamma_\lambda$ may be 
very different from one another due to reordering processes taking place 
in the system at strong coupling to the environment 
(full opening of the quantum billiard). In this regime, 
short-lived and long-lived resonance states coexist  \cite{rep}.
Examples are short-lived bouncing ball modes or whispering gallery 
modes that may coexist with long-lived resonance 
states in a small quantum billiard \cite{naz}.

Whispering gallery modes 
may appear in small chaotic as well as  regular billiards with 
convex boundary when 
fully opened {\it and} the leads are attached to them in a suitable 
manner. Examples are  billiards of Bunimovich  and circular type 
\cite{naz}. The whispering gallery modes are localized  near to the 
boundary of the billiard.
They have an approximately equal distance in momentum $k$ 
from one another, and their positions in energy 
are determined by the number of nodes which 
increases with increasing energy.  
The widths are proportional to the length of the pathway along the 
convex boundary (except for threshold effects) \cite{naz}. 
A shot-noise analysis  has shown 
that they support direct transport processes  \cite{shot}.
They determine, therefore, the optical $S$ matrix.
The long-lived states however feature  indeterministic processes
corresponding to the universal prediction of random matrix theory \cite{shot}.
They cause the fluctuations of the transmission probability.

\subsection{Numerical simulation}

In this section, we show the results of numerical simulations for 
the transmission through a cavity of Bunimovich type with leads attached
in four different ways. The calculations for the transmission are
performed by using the boundary element method \cite{riddell}.
The eigenvalues of $H_{\rm eff}$ are obtained by
applying the tight-binding lattice Green function method  
given in Ref. \cite{datta} and
using the general relation between $H_{\rm eff}$ and the Green function.
The energies and widths of the resonance states are obtained by solving the
fixed-point equations (\ref{fixed1}) and   (\ref{fixed2}).
The ensemble average $\langle t \rangle$ is performed from 200 different 
positions of an
internal obstacle by keeping fixed the area of the
billiard. The energy interval considered is divided usually into 20
energy bins.

\begin{figure}[h]
\epsfig{file=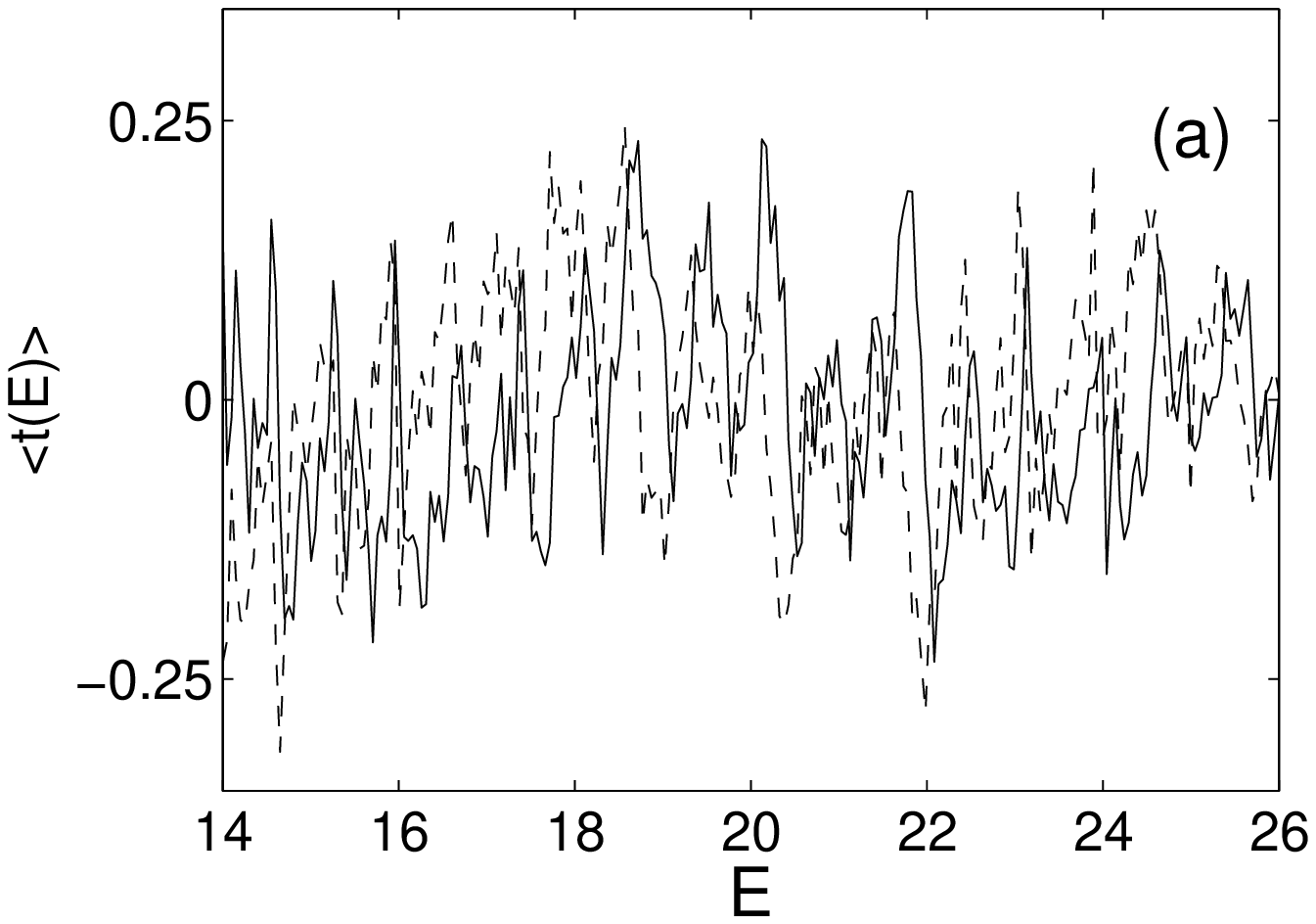,width=7cm,clip=}
\epsfig{file=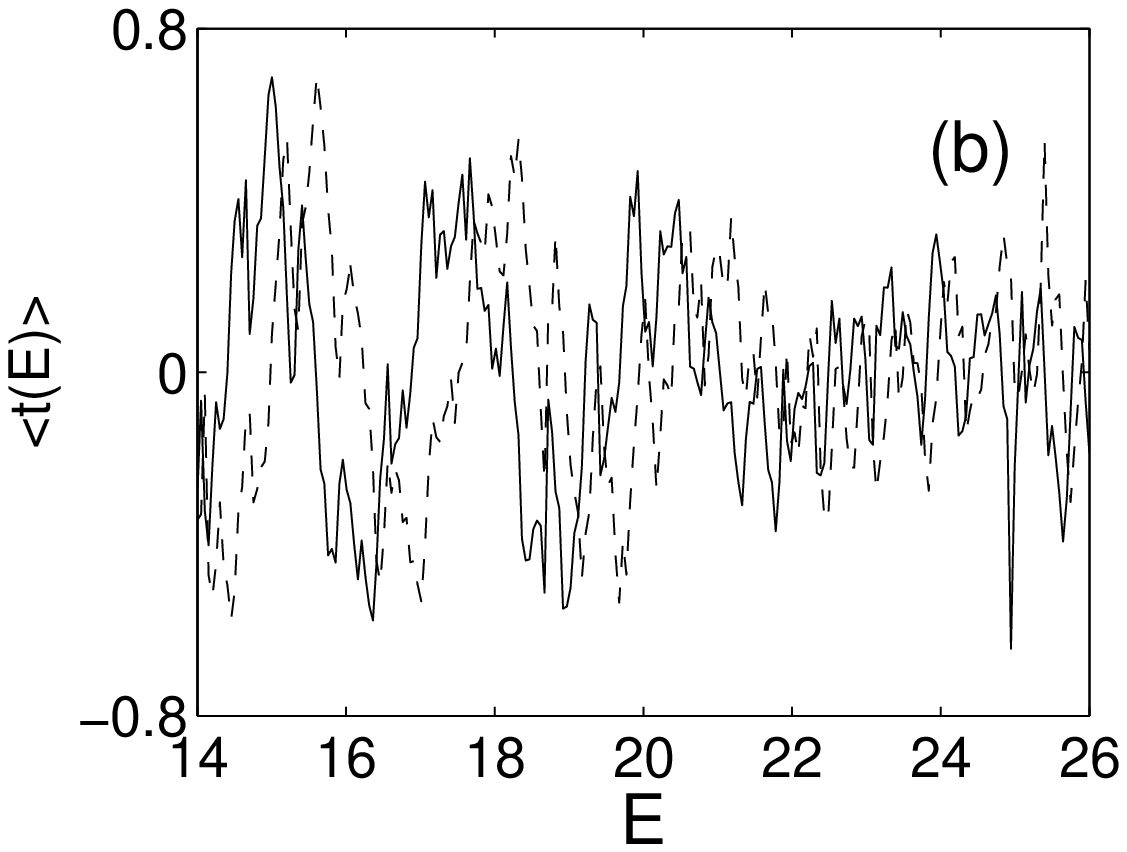,width=7cm,clip=}
\epsfig{file=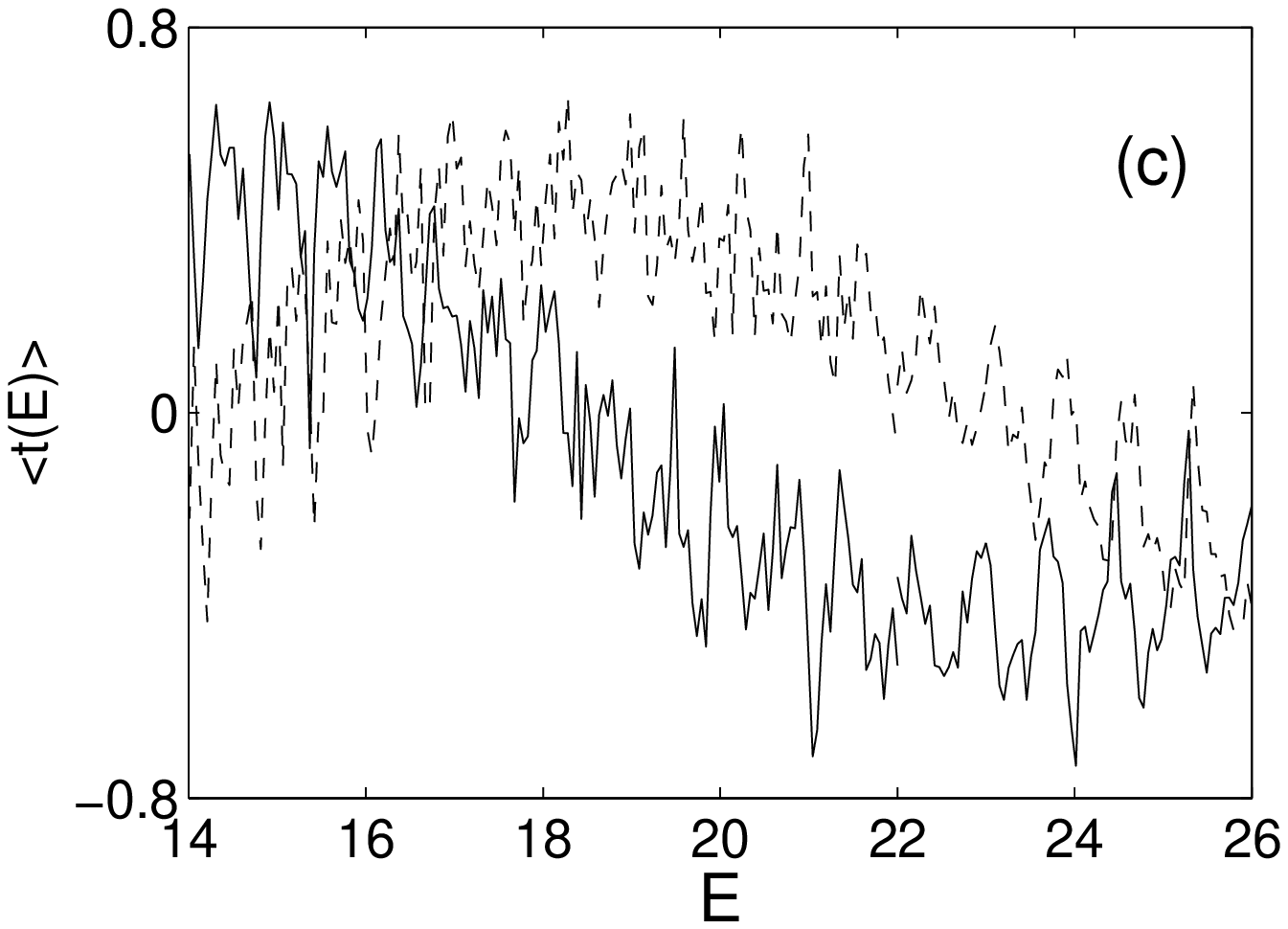,width=7cm,clip=}
\epsfig{file=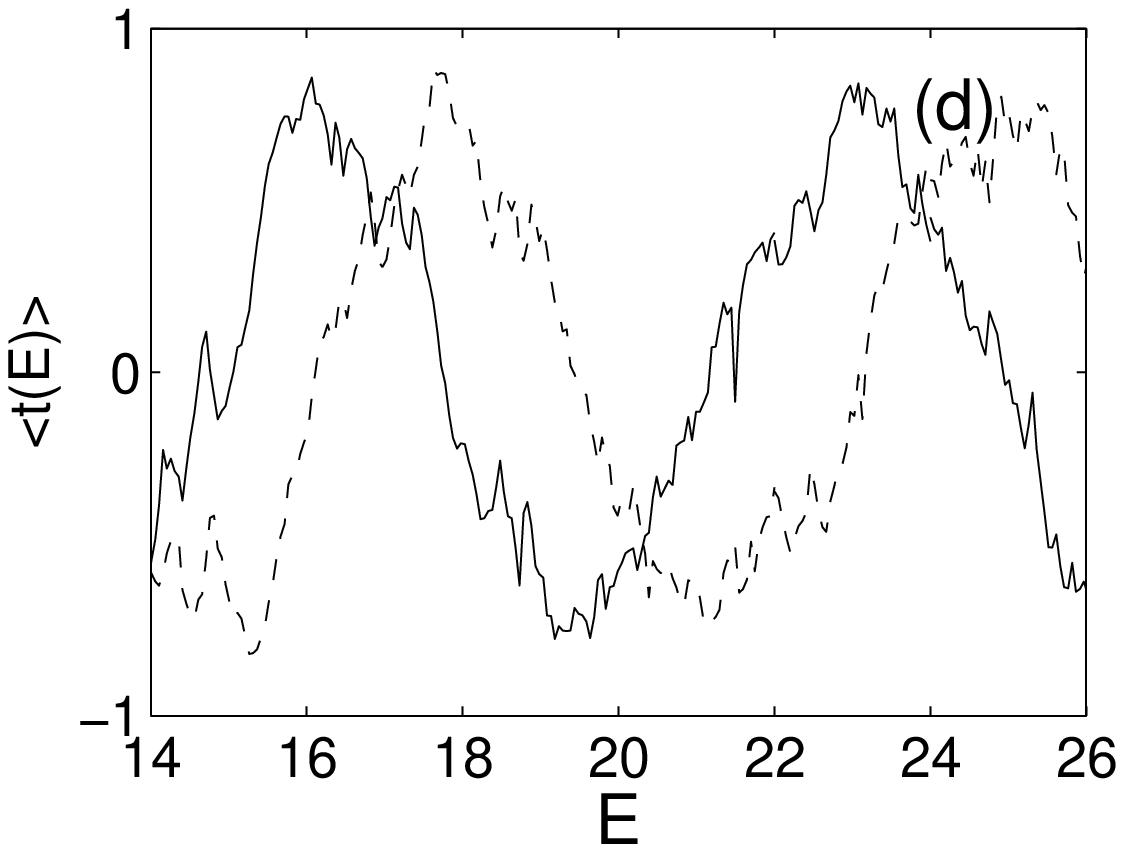,width=7cm,clip=}
\caption{{\small
The ensemble expectation value $\langle t(E)\rangle$ as a function of the
energy $E$ for the cavities (a), (b), (c) and (d) shown in the insets of Fig. 
\ref{fig2}. 
The cavities consist of a Bunimovich stadium connected to two waveguides 
directly, as in panels (a), (b) and (c), or through a smaller half stadium, 
as in (d). Full lines: Re$\langle t(E)\rangle$, dashed lines:  
Im$\langle t(E)\rangle$. Most  $\langle t(E)\rangle$ show oscillations
that are related to the distances between the input and output leads.
The energy is in units $\hbar^2/2m$ and $d=1$ is the width of the
wave guide.
}}
\label{fig1}
\end{figure}

\begin{figure}[h]
\epsfig{file=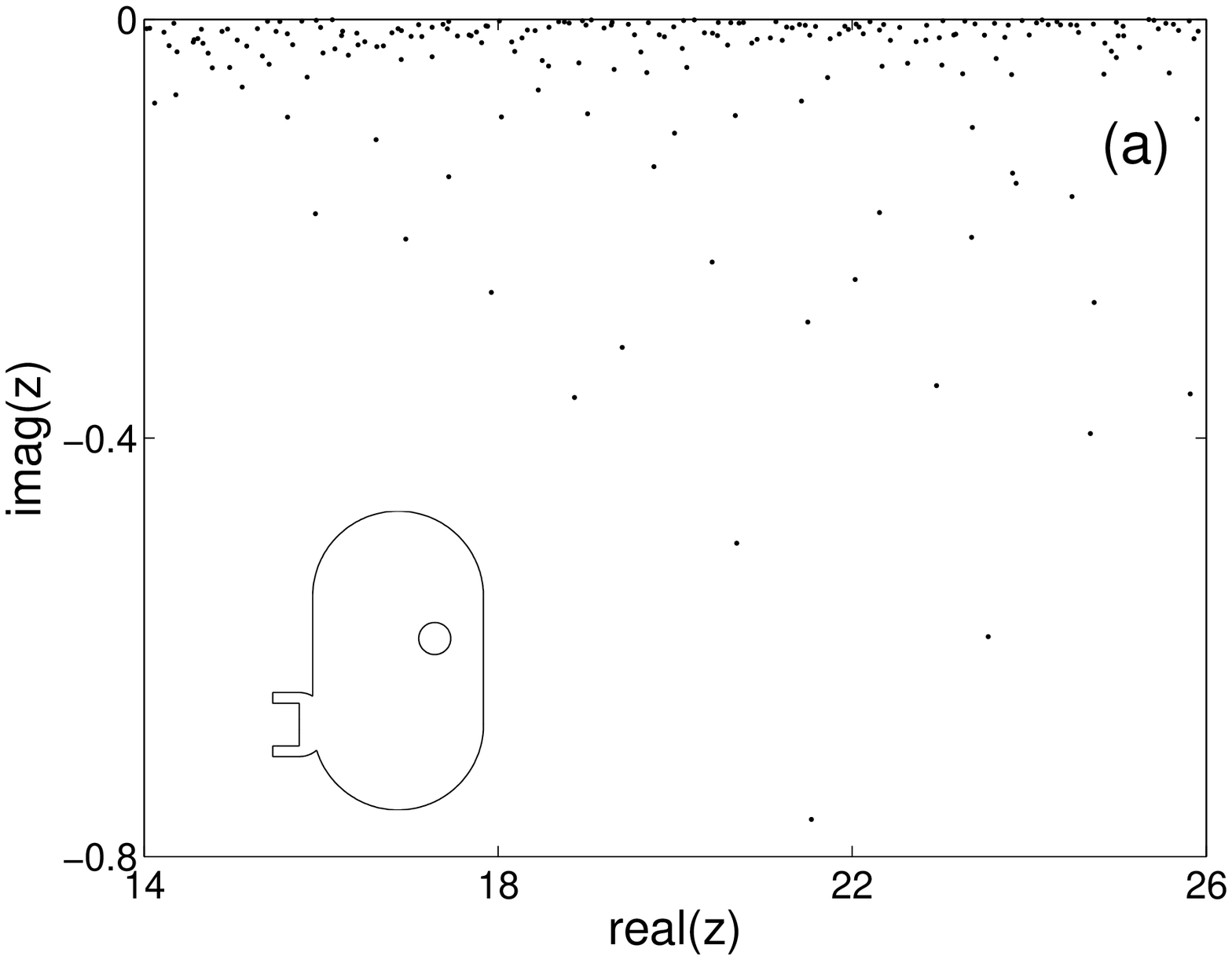,width=7cm,clip=}
\epsfig{file=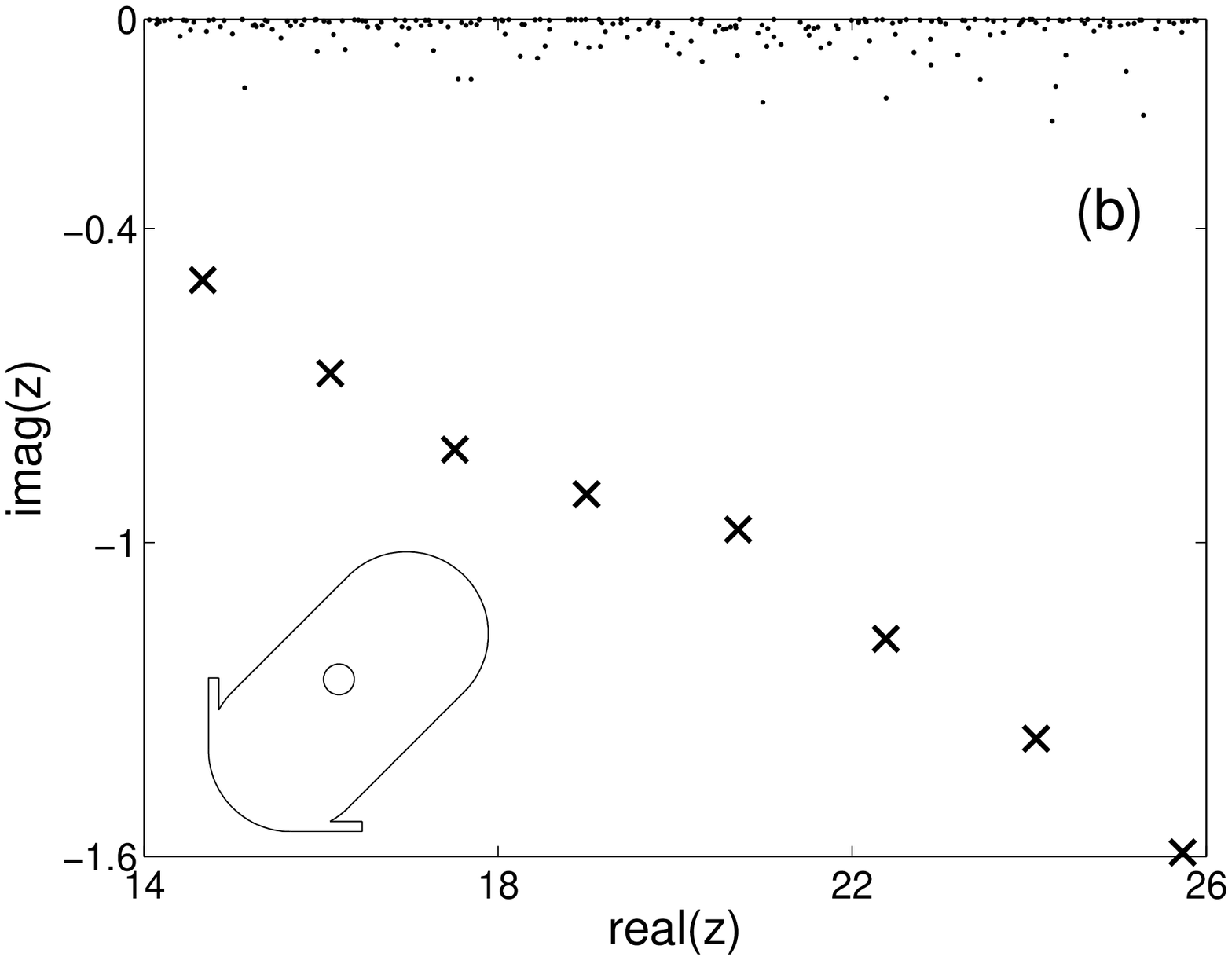,width=7cm,clip=}
\epsfig{file=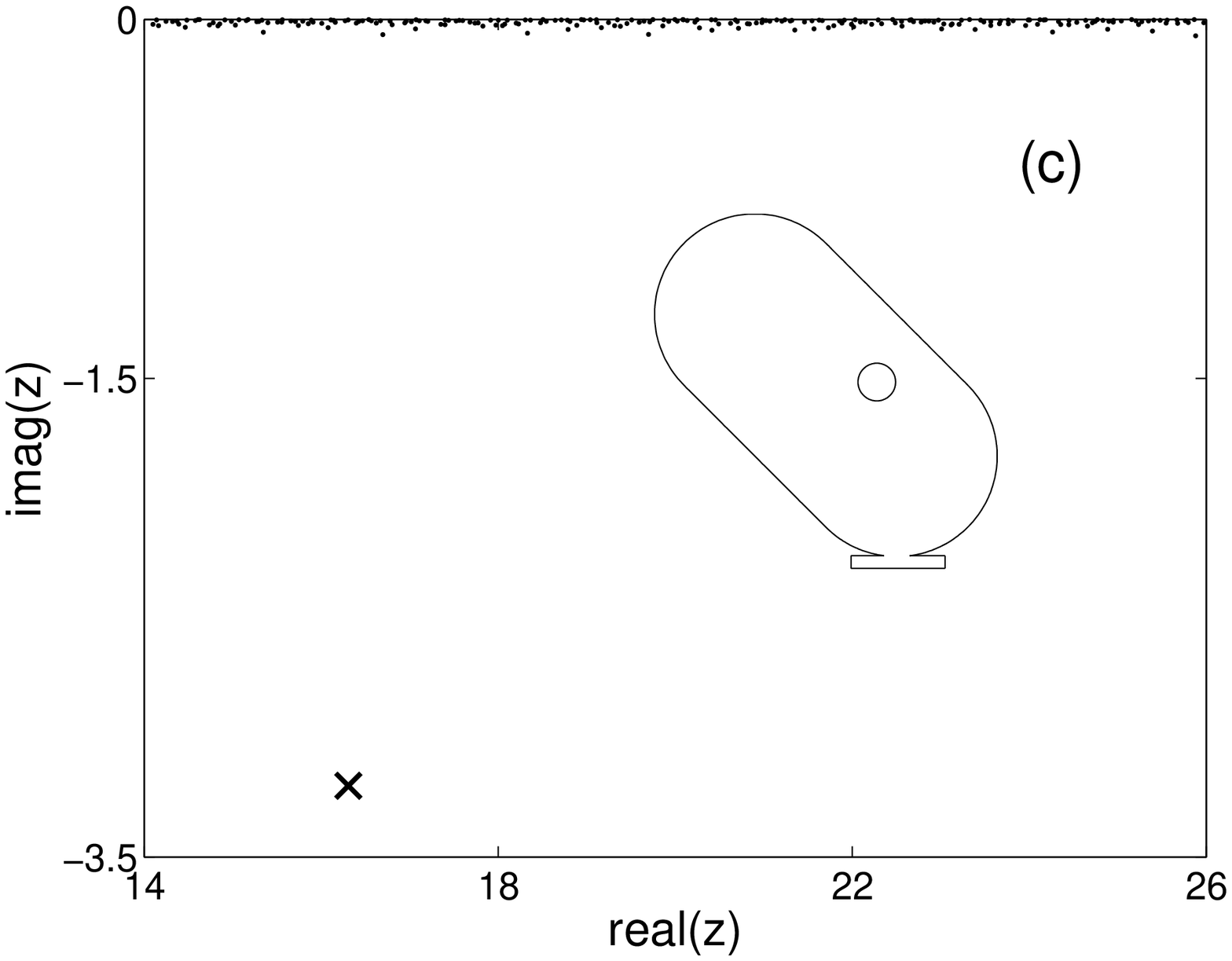,width=7cm,clip=}
\epsfig{file=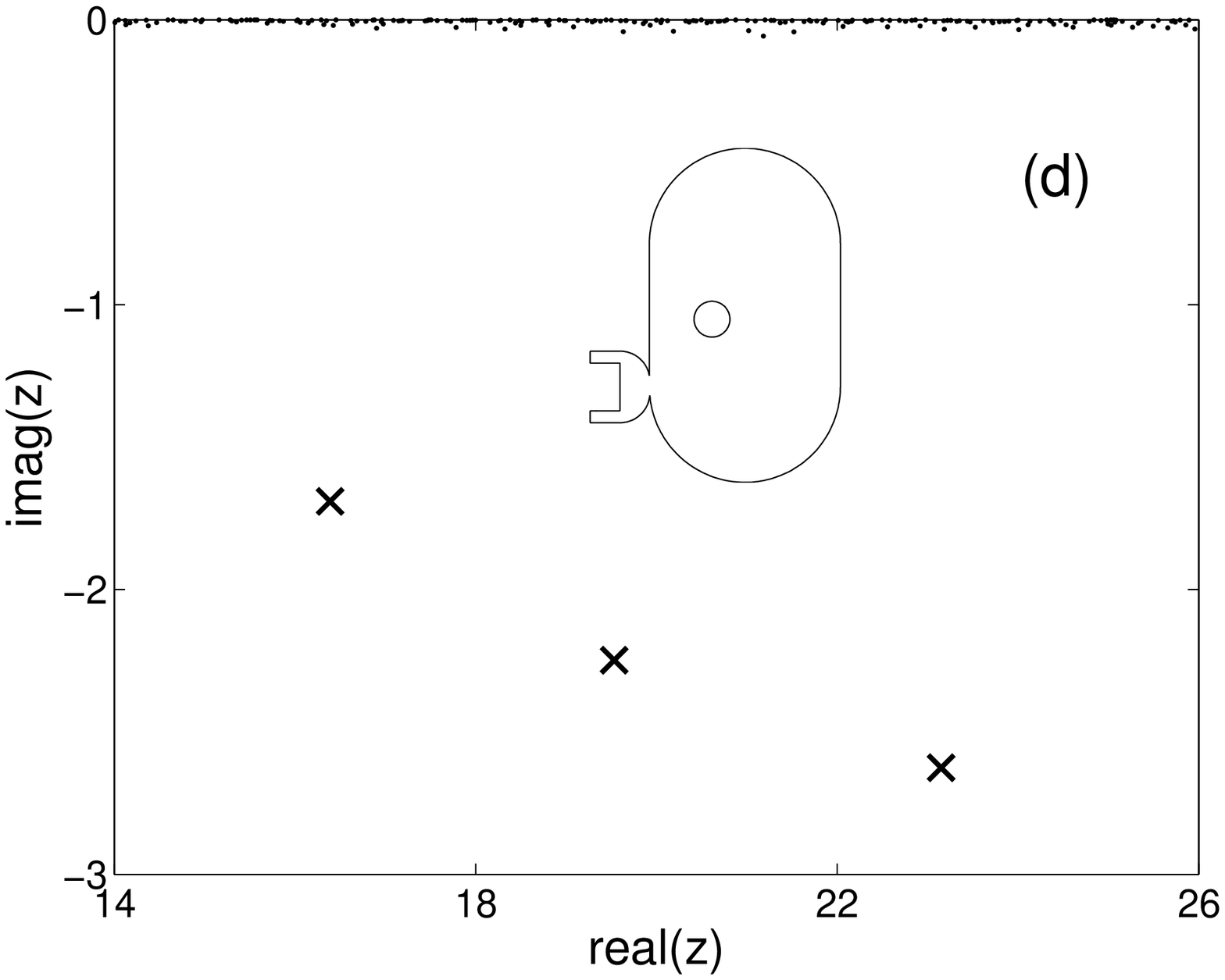,width=7cm,clip=}
\caption{{\small 
The solutions of the fixed-point equations  (\ref{fixed1}) 
and  (\ref{fixed2}) for the 
resonance states of the four different cavities.
The short-lived states are marked by crosses, the long-lived ones by dots.
A clear separation of the time scales can be seen in (c) and (d). 
The neighboring short-lived resonance state in (c) lies at
$E_\lambda - i/2 ~\Gamma_\lambda = 27.74-6.89~i$.
In the insets, the cavities are shown. In order to see the differences 
between  the four open cavities, 
the attached leads are also shown up to an arbitrary  distance $L$. 
The eigenvalues are calculated with $L=0$.
}}
\label{fig2}
\end{figure}

In Fig. \ref{fig1}, we show the ensemble expectation values 
$\langle t(E)\rangle$ 
of the transmission as a function of the energy $E$ in the region  $E=14- 26$
for the four  cavities shown in the insets of Fig. \ref{fig2}. 
The oscillating contribution from the short-lived states 
as well as the shifts between Re$\langle t(E)\rangle$
and Im$\langle t(E)\rangle$ can be seen clearly 
in the cavities  (c) and (d). Interesting is the geometry of the cavity 
(b) where the transmission $\langle t(E)\rangle$ changes its nature at
$E\approx 21$.

In Fig. \ref{fig2}, we show
the eigenvalues of the effective Hamiltonian (\ref{Heff}) 
for the four cavities. The values $E_\lambda$ and $\Gamma_\lambda$
of the short-lived states 
are calculated by solving the fixed-point equations (\ref{fixed1})
and  (\ref{fixed2}).
We have resonance states with well separated time scales in Figs.
\ref{fig2}(c) and (d).
In these cases, $\langle t(E)\rangle$ is large and oscillates.
In Fig. \ref{fig2}(a), separated time scales can not be identified
and $\langle t(E)\rangle$ is relatively small.
In Fig. \ref{fig2}(b), separated 
time scales can be identified but the difference between the short-lived and
the long-lived states is smaller than in (c) and (d).  Furthermore, the
widths of the long-lived states are spread in (b) over a comparably large 
range and the widths of the short-lived states show an irregularity  
around $E\approx 21$. At this energy, $\langle t(E)\rangle$ changes its nature
as can be seen from Fig. \ref{fig1} (b).
In any case, Figs. \ref{fig1}(b) and \ref{fig2}(b) show that the 
sensitivity of $\langle t(E)\rangle$ against parameter
variations is large when the direct pathway between the input and output 
leads is large and not well separated from other
pathways through the interior of the cavity. This sensitivity can be seen
also in comparing the results for the cavities (a) and (d). While there is
almost no separation of the pathways through
the small attached half stadium from those through the large
Bunimovich stadium in (a), both parts are well separated in (d). 
As a consequence,
we see whispering gallery modes in the small attached stadium in (d) but not
in (a).

The short-lived states determine the value of the so-called optical 
$S$ matrix.


\section{Eigenfunctions of the effective Hamiltonian and
phase rigidity }
\label{rigid}

\subsection{Eigenfunctions and transmission}

\begin{figure}[h]
\epsfig{file=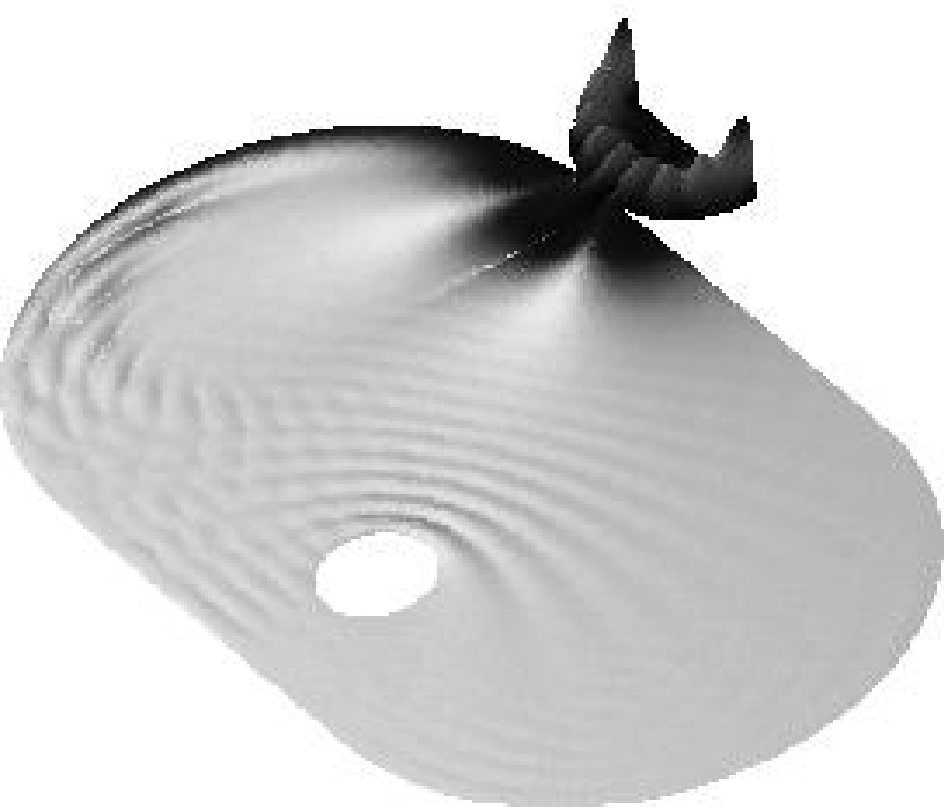,width=6cm,clip=}\hspace*{1.3cm}
\epsfig{file=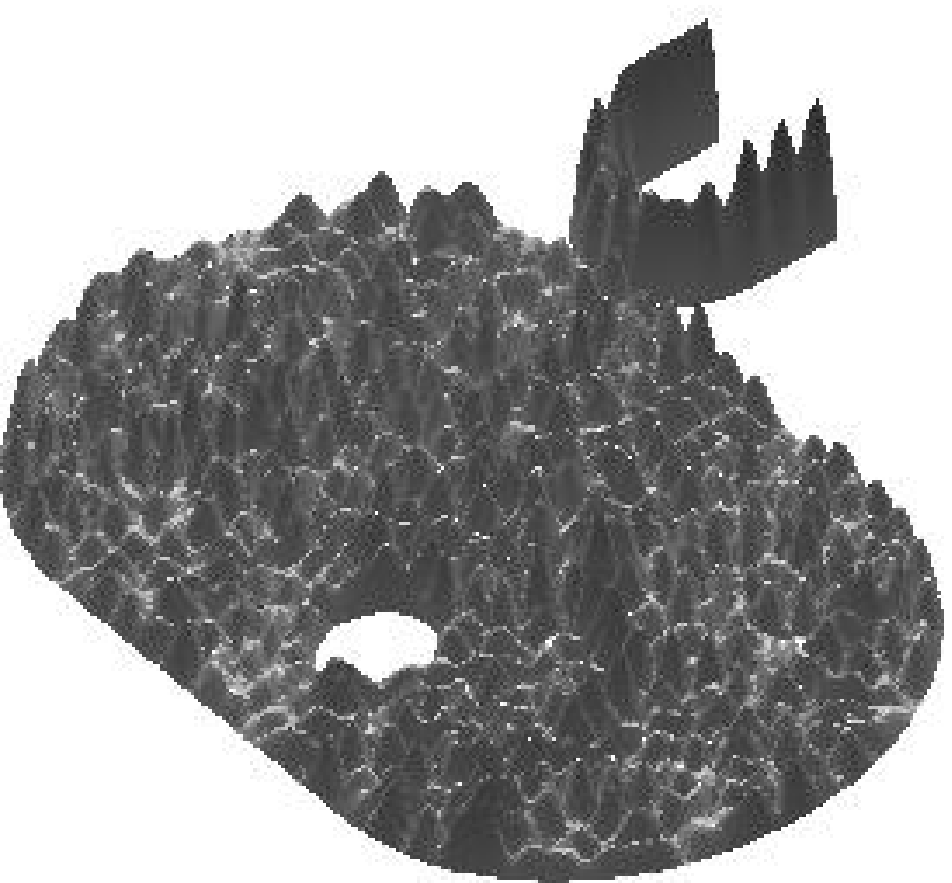,width=6cm,clip=}
\caption{{\small 
The eigenfunction $\phi_\lambda$ of the effective Hamiltonian $H_{\rm eff}$
at the energy $E=E_\lambda$ of the whispering gallery mode $\lambda$ 
(left) and the scattering wave function
$\Psi^E_C$  at the same energy $E$ (right). 
$E_\lambda - i/2 ~\Gamma_\lambda = 19.53-2.25~i$.
The $\phi_\lambda$ is shown up to the attached
lead ($L=0$) while the $\Psi^E_C$ is shown, for illustration, also in the lead
up to an arbitrary  finite value $L$.
}}
\label{fig3}
\end{figure}

The scattering wave function $\Psi_C^E$ is solution of the
Schr\"odinger equation $(H-E)\Psi^E_C=0$ in the
total function space with the hermitian Hamilton operator $H$. 
It reads \cite{rep,saro}
\begin{equation}
\label{total}
\Psi_C^E= \xi^E_C + \sum_{\lambda}\bigg[ \phi_\lambda \,+ \,
\xi^E_C \, \frac{1}{E^+-H_C} \, \langle \xi^E_C|V|\phi_\lambda) \bigg] \,
\frac{(\phi_\lambda |V| \xi^E_C\rangle}{E-z_\lambda} \, . 
\end{equation}
The $\phi_\lambda$ are complex and  biorthogonal \cite{rep}, 
\begin{equation}
(\phi_\lambda|\phi_{\lambda '}) \equiv \langle\phi_\lambda^*|\phi_{\lambda '}
\rangle = \delta_{\lambda, \lambda '} 
\label{biorth1}
\end{equation}
\begin{equation}
|\langle\phi_\lambda|\phi_{\lambda}\rangle | = A_\lambda \ge 1 \; ; \qquad
|\langle\phi_\lambda|\phi_{\lambda '}\rangle | = B_\lambda^{\lambda '} \ge 0
\; . 
\label{biorth2}
\end{equation}
Eq. (\ref{total}) shows that the
scattering wave function $\Psi^E_C$ in the interior region 
of the quantum dot is determined, above all, by the complex eigenfunctions
$\phi_\lambda$ of $H_{\rm eff}$,
\begin{eqnarray}
\Psi_C^E(r)\to \sum_\lambda  \phi_\lambda (r) \,
\frac{(\phi_\lambda |V| \xi^E_C\rangle}{E-z_\lambda} \; .
\label{total1}
\end{eqnarray}
At the energy $E\approx  E_\lambda$, the eigenfunction $\phi_\lambda(r)$ of 
the effective Hamiltonian $H_{\rm eff}$ gives the main contribution.
This holds true especially at the energy of a narrow resonance state.
As a numerical example, the eigenfunction $\phi_\lambda$ of a whispering
gallery mode is shown in Fig. \ref{fig3} for the cavity (d)
and compared with the 
corresponding scattering wave function $\Psi_C^E$ 
at the energy $E_\lambda$. The scattering wave function contains
contributions from  other eigenfunctions also at $E=E_\lambda$.
Nevertheless, it shows, at this energy, the typical structure  
of the whispering gallery mode. As can be seen from the figure, it is
localized not inside the large Bunimovich cavity but outside of it
in the small half stadium.

The real and imaginary parts of the eigenfunctions $\phi_\lambda$ are more or
less decoupled in the regime of overlapping resonances \cite{rstopol}. 
The value
\begin{eqnarray}
r_\lambda & = & 
\bigg| \frac{\int dr (|{\rm Re}\, \phi_\lambda (r)|^2 - 
|{\rm Im}\, \phi_\lambda (r)|)^2}
{\int dr (|{\rm Re}\, \phi_\lambda (r)|^2 + |{\rm Im}\, \phi_\lambda (r)|)^2}
\bigg| 
= \bigg| \frac{(\phi_\lambda | \phi_\lambda)}{\langle
\phi_\lambda | \phi_\lambda \rangle}  \bigg|
=  \frac{1}{ A_\lambda}
\label{rigbio}
\end{eqnarray}
is a measure for the biorthogonality of the eigenfunctions of $H_{\rm eff}$. 
The phase rigidity $|\rho|^2$ of the scattering wave function $\Psi(r)$  
is considered in e.g.  Refs. \cite{brouwer1,brouwer2},
\begin{eqnarray} 
\rho =\frac{\int dr ~\Psi(r)^2}{\int dr ~|\Psi(r)|^2}
= e^{2i\theta} \frac{\int dr ~(|{\rm Re}\tilde\Psi(r)|^2 - 
|{\rm Im}\tilde\Psi(r)|^2)}{
\int dr ~(|{\rm Re}\tilde\Psi(r)|^2 + |{\rm Im}\tilde\Psi(r)|^2)} 
\label{rig}
\end{eqnarray}
where $\theta$ is an angle providing that Re$\tilde\Psi(r)$ and  
Im$\tilde\Psi(r)$ are orthogonal. The value $\rho$
is related to the $r_\lambda$ according to (\ref{total1}).

For an isolated 
resonance state, $A_\lambda \approx 1$ and $r_\lambda  \approx 1$ 
at the energy  $E=
E_\lambda$. At this energy, the transmission probability shows a peak.
Approaching a branch point in the complex energy plane \cite{rep} where two 
eigenvalues $z_{\lambda_1}$ and $z_{\lambda_2}$ coalesce,  
$A_\lambda\to \infty$ and $r_\lambda \to 0$ \cite{ro01,rstopol}. Here, 
the widths bifurcate~: one of the states aligns with the channel wave 
function and becomes short-lived while the other one becomes long-lived
\cite{rep}. Eventually, the short-lived and long-lived resonance 
states differ strongly from one another and do not cross in the complex energy
plane \cite{rstopol}. 
Therefore again, as for non-overlapping resonances,
$A_\lambda \to 1$ and $r_\lambda \to 1$  
at the energies  $E= E_\lambda $ of the long-lived states.
In the transmission through the cavity,  the short-lived states 
determine the smooth 'background' while the long-lived states cause 
the superposed peaks (fluctuations) 
in the transmission probability \cite{rstrans}. 

This picture can be translated to that of 'standing' and 'traveling' waves.
Standing waves 
\begin{eqnarray} 
\psi (r) = (2N)^{-1/2} \sum_{n=1}^{N} {\rm cos} ~[(\theta_n +k_n\cdot r)] 
\label{stan1}
\end{eqnarray}
cause  the Porter-Thomas statistics for the intensity \cite{shapiro}. 
Tuning the frequency to a resonance, an intensity pattern 
is generated that closely follows the profile of this resonance.
In this situation, the resonances are isolated from one
another and $A_\lambda \to 1$, $r_\lambda \to 1$. 

When the cavity is fully open,  the local field  can be viewed as a
sum of a number of traveling modes ariving at a point from various
scattering processes \cite{shapiro},
\begin{eqnarray}
\psi (r) = (2N)^{-1/2} \sum_{n=1}^{N} {\rm exp} ~[i(\theta_n +k_n\cdot r)]
\label{trav1}
\end{eqnarray}
where the phases $\theta_n$ are completely random and the wave vectors $k_n$
are uniformly distributed.
Both Re$(\psi)$ and Im$(\psi)$ are independent Gaussian
variables what leads to the Rayleigh distribution for the  intensity 
$I(r)\equiv |\psi(r)|^2$. 
It applies to a monochromatic wave propagating in an open system 
('traveling wave' excited by a monochromatic source \cite{shapiro}).
In the corresponding description of this situation
with the effective Hamiltonian formalism,
short-lived and long-lived resonance states coexist in the
system and determine, respectively, the smooth 'background'  and
the superposed peaks (fluctuations) of the transmission probability
\cite{rstrans}. The short-lived
resonance states are strongly related to the channel wave functions 
(scattering states in the leads)
due to the large overlap integral of their wave functions with those of
the channel wave functions. The  transmission 
induced by these states shows therefore the same dependence on the
momentum $k$ as the scattering wave functions $\xi^E_C$ in the leads
('traveling' waves). Obviously, the 'monochromatic source' by which the 
traveling wave is excited according to \cite{shapiro} is,
in the one-channel case,  the channel wave 
function $\xi^E_C$ since it causes the alignment of one of the wave functions
$\phi_\lambda$ in approaching
the branch point in the complex energy plane \cite{rep}. 
The traveling waves determine the optical $S$ matrix.

\subsection{Numerical simulation}

\begin{figure}[h]
\epsfig{file=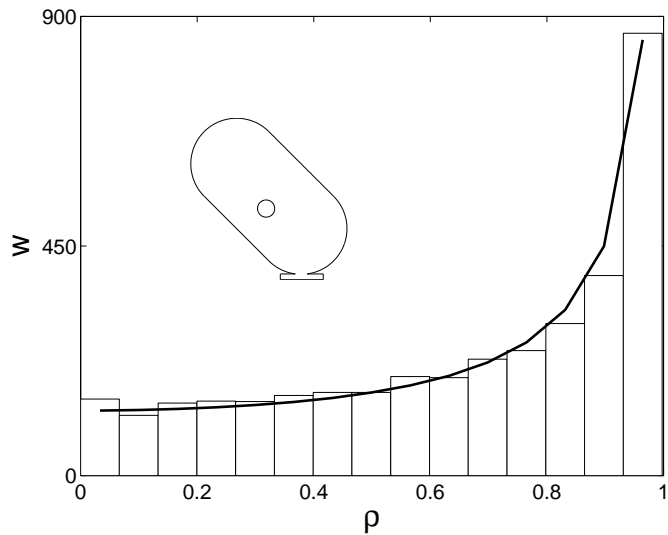,width=9cm,clip=}
\caption{{\small 
The ensemble and energy averaged  phase rigidity  (\ref{rig}) for  
resonance states of the  cavity (c). Energy interval [22, 23].
The full line is calculated from the equation for the phase rigidity
distribution in the case of a chaotic cavity with energy 
averaging and 2 channels \cite{brouwer2}.
The results for the cavities (a), (b) and (d), see insets in Fig. \ref{fig2},
in the same energy interval [22, 23] as well as those for the cavity
(b) in the energy interval [20.5, 21.5] are nearly the same.
}}
\label{fig4}
\end{figure}

The distribution of the phase rigidity is calculated by means of (\ref{rig}).
We take the expectation value $\langle \,|\rho|^2\rangle$ 
for an ensemble of 200 cavities with different positions of the obstacle in
the interior and for 20 different energy  values inside the energy interval
considered. The results are almost the same in all cases considered, i.e. 
for the four different open cavities in the energy interval [22, 23] and,  
in addition, for the cavity (b) in the
energy interval [20.5, 21.5].  A typical result is shown in Fig. \ref{fig4}. 
It agrees well with the theoretical value (full line in Fig. \ref{fig4}) 
for a chaotic cavity and
2 channels \cite{brouwer2}. The phase rigidity is mostly near to its 
maximum value in the two-channel case. 
This corresponds to the fact that the transmission is caused by either
standing waves or traveling waves with superposed fluctuations. 
The transition between the two scenarios takes
place in a comparably small region according to the examples
studied in \cite{rstrans,jung}.


\section{Summary}

The spectral properties of an open cavity depend strongly on the manner the
leads are attached to it. We studied the eigenvalues and eigenfunctions
of the effective Hamiltonian $H_{\rm eff}$ describing a large cavity 
of Bunimovich type with two leads attached in four different ways
and one channel in each lead. 
The transmission is resonant in all cases in relation to the effective
Hamiltonian of the open quantum system.
In some cases, the transmission takes place via standing waves in the cavity
with an intensity that  closely follows the profile of the resonances.
In other cases, two different types of resonance states appear
which differ by their lifetimes. The
short-lived  states cause traveling modes  while the
long-lived states appear as fluctuations of the transmission probability.
The short-lived states, including the whispering gallery modes, can 
{\it not} be identified in the interior of 
large cavities where the pathway between input and output 
leads is large. Therefore their widths  are relatively small in
this case, and it is impossible to identify them in the 'sea' of long-lived
resonance states.

The optical $S$ matrix is related to the short-lived states 
(traveling waves) as can be seen
from the eigenvalues and eigenfunctions of $H_{\rm eff}$.
This relation is however not necessarily true also in the opposite direction~:
the eigenvalues of $H_{\rm eff}$ may show separated time scales while the
optical $S$ matrix is, nevertheless, small. In such a case, the short-lived
modes exist, according to our numerical results,  inside the cavity, 
and their spectroscopic properties are very
sensitive against small parameter changes. In the four cases studied by us,
ballistic modes do not appear in the interior of the large chaotic cavity.

In all considered cases, the phase rigidity fluctuates  as a 
function of energy but is mostly near to its maximum value.
The distribution is characteristic of the two-channel case.


\acknowledgments

We are indebted to  Victor Gopar and Pier Mello for the many discussions 
on the different
possibilities to attach the leads to the Bunimovich stadium.
We thank T. Gorin for a critical reading of the manuscript.
E.N.B. is grateful to the MPI-PKS for its hospitality during his
stay in Dresden. This work has been supported by RFBR grant 05-02-97713
"Enisey".


\end{document}